\documentclass[twocolumn]{aa}
\usepackage{graphicx}
\usepackage{txfonts}
\usepackage{amssymb}
\usepackage{times}
\usepackage{natbib}
\bibpunct{(}{)}{;}{a}{}{,}
\usepackage{color}
\usepackage{xspace}
\usepackage{ulem}

\def\DS{D_\mathrm{S}}
\def\DL{D_\mathrm{L}}

\def\ThE{\theta_\mathrm{E}}

\def\rhoS{\rho_\mathrm{*}}
\def\tE{{t_{\rm E}}}
\def\to{{t_{\rm 0}}}
\def\uo{{u_{\rm 0}}}
\def\x{{{z}}}
\def\y{{{\zeta}}}

\def\yent{{\zeta_{\rm entry}}}
\def\yexi{{\zeta_{\rm exit}}}
\def\tent{{t_{\rm entry}}}
\def\texi{{t_{\rm exit}}}
\def\sent{{s_{\rm entry}}}
\def\sexi{{s_{\rm exit}}}
\def\K{{\mathrm{cst}}}
\def\Ac{{A_{\rm C}}}
\def\Anc{{A_{\rm NC}}}
\def\yperp{{\y_\perp}}
\def\xc{{\x_c}}
\def\yc{{\y_c}}
\def\tc{{t_c}}

\def\mTer{M_{\oplus}}

\newcommand\Eq[1]{Eq.~(\ref{#1})}
\newcommand\Fig[1]{Fig.~\ref{#1}}

\newcommand\Sec[1]{Sec.~\ref{#1}}
\def\cf{{\it cf.~}}
\def\d{\mathrm{d}} 
\newcommand\dix[1]{\times 10^{#1}}

\usepackage{ulem}
\usepackage{color}

\begin{document}

\title{An alternative parameterisation for binary-lens caustic-crossing
  events}  
\titlerunning{An alternative parameterisation for binary-lens
  caustic-crossing events}  
\authorrunning{A.~Cassan}
\author{A.~Cassan}
\institute{
  {Astronomisches Rechen-Institut (ARI), Zentrum f\"{u}r Astronomie
    (ZAH), Heidelberg University, M\"{o}nchhofstra{\ss}e 12--14, 69120
    Heidelberg, Germany}  } 
\date{Received ; accepted}
\abstract
{}
{ Microlensing events are being discovered and alerted by the two survey
  teams OGLE and MOA at an increasing rate.
  Around ten percent of these events involve binary lenses. Such
  events potentially contain much information on the physical
  properties of the observed binary systems, which can then be used for
  e.g. statistical studies on binary objects in the Galactic
  disk or bulge. However, such events are usually not straightforward
  to study, 
  because the model equations are strongly non-linear and there are
  many local minima that can fool the search for the best solution if
  the parameter space is not inspected with great care. In this work 
  an alternative parameterisation for the binary lens fitting
  problem is proposed, in which the parameters involved are defined to represent
  as closely as possible the caustic-crossing features observed in most
  binary lens light curves. Furthermore, we work out an extension of
  the method in order to make use of the straight line fold caustic
  approximation, when the latter applies for both the caustic entry and
  exit. } 
{ We introduce an alternative parameterisation in order to
  confine the exploration of the parameter space to regions where the
  models only involve caustic crossing at the dates seen in the
  light curve.} 
{ We find that the proposed parameterisation provides more robustness to
  the light curve fitting process, in particular in avoiding a code to
  get stuck in false minima. } 
{}

\maketitle

\section{Introduction}

Since the original idea of \cite{Paczynski1986} to use the
gravitational lensing effect to track the presence of dark objects
in the Milky Way, more than three
thousand microlensing events have been 
discovered, mostly in the
direction of the Galactic centre. A microlensing event occurs
when the light from a background bright star (or
source) is bent by the gravitational field of an intervening
massive body (or lens) passing close to the observer line-of-sight. 
The phenomenon produces an apparent transient brightening of the
source flux, resulting in a typical bell-shaped light curve. 
Current surveys operated by the OGLE \citep{Ref-OGLE}
and MOA \citep{Ref-MOA} collaborations provide lists of ongoing
microlensing events by monitoring more than $10^{8}$
stars towards the Galactic bulge on a daily basis, 
while (up to the 2007 bulge season) collaborations 
like PLANET, RoboNet or $\mu$FUN use ground-based
networks of telescopes to perform a round-the-clock follow-up of
selected events.

Since the information on the lens characteristics 
is contained in the lens mass
distribution rather than in its luminosity, microlensing naturally
opens a new window in the exploration of invisible objects within the
Galaxy. Indeed, shortly after \cite{Paczynski1986}'s work,
\cite{MaoPaczynski1991} proposed microlensing as a promising tool to
hunt for extrasolar planets, a few years before the first planets were
discovered around a pulsar \citep{WolszczanFrail1992} by pulse timing,
and around a solar-like star \citep{MayorQueloz95} by radial
velocity measurements. 

Searching for planets is today the leading motivation of Galactic
microlensing observations. Amongst the $\sim 300$ 
exoplanets discovered so far, most of them using the radial velocity
technique, microlensing contributes seven planets. Four of them are 
giant planets of a few Jupiter masses: MOA~2003-BLG-53/OGLE~2005-BLG-235Lb
\citep{OGLE03235Lb}, OGLE~2005-BLG-071Lb
\citep{OB05071Lb,Dong2008ob71}, plus a slightly less massive
Saturn/Jupiter planetary system analogue 
\citep[OGLE~2007-BLG-109Lb+c,][]{OGLE07109Lbc}, orbiting low-mass
dwarf stars. A similar number of super-Earth planets have been found
as well: OGLE~2005-BLG-390Lb \citep{OGLE05390Lb,Jovi2008}, a $5.5$
Earth-mass planet at $\sim 2.6$ AUs which makes it the first ever
rocky/icy planet discovered around a main sequence host beyond the snow
line, OGLE~2005-BLG-169Lb
\citep{OB05169Lb}, $13$ times more massive than Earth, and recently
MOA~2007-BLG-192Lb \citep{MOA192Lb}, one of the least
massive planet discovered so far with an estimated mass of only $\sim 3\mTer$,
probably orbiting a very late M or brown dwarf. 

While microlensing has so far contributed little to the large number of
extrasolar planet discoveries, it probes a region in the mass/orbit
diagram -- namely cool low-mass planets (cold Neptunes and super-Earths)
at few AUs from their stars -- where no other competitive technique has
the required sensitivity. Furthermore, microlensing is
sensitive  to extremely low mass bodies, so that planets of mass
below that of the Earth could be reached by a dedicated space mission
\citep{BennettRhie2002}.   

Most of the microlensing events detected do not show evidence
of the presence of planets around the lenses.
For these so-called single lens events, 
modelling is relatively easy and it is possible to
use a large number of events to draw statistical 
conclusions on the abundance of
planets. Such studies have been carried out in particular by
\cite{Gaudi5years} for a selection of PLANET events in the period
1995--1999, \cite{Tsapras2003} with a sample based on OGLE events
between 1998--2000, and \cite{Snodgrass2004} using data form the
OGLE 2002 campaign.
Nevertheless, the nature of many anomalous events remains unclear
because their modelling requires much more work and computational
resources. Up to now, such events have led to separate studies in order to
derive constraints on the presence of 
planets around the lens, as e.g. performed for the very high
magnification event ({\it i.e} very sensitive to planetary signals)
OGLE~2007-BLG-050 \citep{BatistaOB07050}.

The steadily increasing number of anomalous 
microlensing events is thus posing a
great challenge to modellers, for the proper treatment of an individual
case requires one to probe many models, hence demanding very efficient
modelling schemes based on robust mathematical and numerical methods.
A large fraction of binary lens events
become unmodelled each season. Since these represent close to $10\%$
\citep{2004AcA....54..103J} of the overall sample (close to a thousand
events a year), a special effort to  
provide improved and more automated modelling strategies is needed.

Events that exhibit caustic crossings are usually very
rewarding in terms  
of extracting stronger constraints on the physics of the binary lens system,
such as the masses and physical separations of the individual components
\citep[e.g.][]{2005AcA....55..159J,Kubas2005}. 
The first microlensing
event showing observational evidence of a 
binary lens caustic crossing is OGLE No.~7 \citep{OGLE-n.7}, 
and the first planet discovered
by microlensing MOA~2003-BLG-53/OGLE~2005-BLG-235Lb
\citep{OGLE03235Lb} also involves a caustic crossing, emphasing the
importance of such binary lens events. 

Although the light curve features associated with caustic crossings
can be very pronounced and easily recognised, the underlying best model
can be hard to find since the parameter space often is afflicted by
many local minima, even if the light curve is well covered.
This problem is aggravated by using parameterisations  
in which the trajectory is modelled by parameters
that are not well related to the light curve features.
This is for example the case in the widely used binary lens
standard parameterisation using the angle between the binary 
lens axis and the minimum distance between the trajectory and the origin of the
coordinate system. A main reason for this is that
the equations of lensing are strongly non-linear
and changing a given trajectory parameter can have very unexpected
consequences for the shape of the light curve. This is explained by
the presence of extended gravitational caustics when the lens
is composed of more than one body. When the source crosses a caustic, its flux
experiences a very steep magnification gradient. Hence, even a small change in
the trajectory can cause a very strong change in flux.  
A desired characteristic of an alternative parameterisation would be
to directly relate the model parameters to visible features in the
light curve, such as the caustic crossing dates, whatever the
source trajectory. All probed configurations in the minimisation
process would then display caustic crossings at their observed
position on the light curve. We aim here to provide an alternative
parameterisation of the binary lens caustic crossing model that works
in this way. 

The paper is organised as follows: in \Sec{sec:meth}, we introduce the 
basic principles of the method and describe the new parameterisation, in
particular the choice of a curvilinear abscissa along the caustic
lines. In \Sec{sec:numeric} we detail the implementation of the
method, and discuss the possible fitting strategies and the inclusion
of more elaborate binary lens models within this framework. We also tackle
some important aspects arising from the introduced set of parameters.
In \Sec{sec:linecaus}, we work out in detail a possible way to
implement the so-called straight line fold caustic approximation within the
presented framework, which allows under some assumptions extremely
efficient computations. We summarise in \Sec{sec:conclusion} our main
results, and discuss the broader implication of fitting schemes and
strategies to provide more automated modelling software in the context 
of future generations of robotic telescopes.

\section{The method} \label{sec:meth}

\subsection{Caustic curves} \label{sec:caustics}

The fundamental equation of Galactic microlensing is the lens
equation, which links the angular position of a point-source object
to the angular positions of its multiple images 
in units of $\ThE$, the angular Einstein radius \citep{Einstein1936}:
\begin{equation}
  \ThE = \sqrt{\frac{4GM}{c^2}
    \left(\frac{\DS-\DL}{\DS\,\DL}\right)}\, , 
\end{equation}
with $\DS$, $\DL$ the observer-source and -lens distances respectively,
and $M$ the lens total mass.
In the formalism of \cite{Witt1990} used here, the binary
lens equation is written using the complex notation:
\begin{equation}
  \y = \x - \frac{1}{1+q}\,\left(\frac{1}{\overline\x}
  + \frac{q}{\overline\x+d}\right),  
\label{eq:eqlensprim}
\end{equation}
where we have chosen the coordinate system so that the more massive
component of the lens is located at the origin, with the secondary
body located on the left at a separation\footnote{The term
  ``separation'' has to be understood as the projected distance
  between the two lens components on the plane of the sky, in units
  of $\ThE$.} $d$. 
Finally, $q\leq1$ denotes the binary lens mass ratio. 

\begin{figure}[!htpb]
  \begin{center}
    \includegraphics[width=8.5cm]{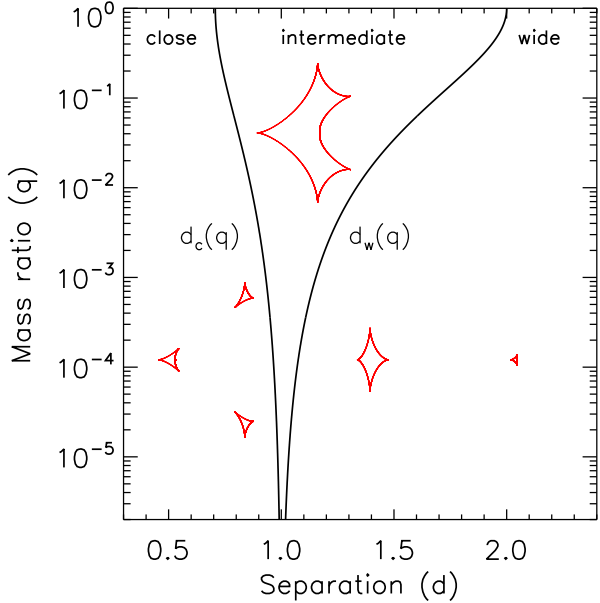} 
    \caption{
      The three topologies of binary lenses: the
      close binary lens (on the left), which involves a central
      caustic plus two off-axis small secondary caustics (plotted here
      for $d=0.8$ and $q=10^{-2}$), the intermediate binary (in the
      middle, assuming $d=1$ and $q=10^{-2}$)  with a single caustic,
      and the wide binary (on the right, $d=1.6$ and
      $q=10^{-2}$) whith a central caustic and an isolated
      secondary caustic. The figure also shows the
      bifurcation functions $d_c(q)$ and $d_w(q)$ between the three
      regimes, which are respectively displayed from left to right as
      the two vertical black lines. When the mass ratio tends to zero,
      the extension of the intermediate domain also tends to zero.}
    \label{fig:caustopo}
  \end{center}
\end{figure}

Multiple lenses in general harbour caustic lines, which are
imaginary lines on the plane of the sky where the magnification of a
perfect point-source diverges. Mathematically, they draw
closed curves where the Jacobian of the transformation
$\x\rightarrow\y$ vanishes. In the following and for brevity, we will
refer to the term ``caustic'' not only for the lines themselves, but
also for the closed curves they form. 
In the case of binary lenses, caustics exist in three different
topologies, usually referred as {\it close}, {\it intermediate} and {\it wide}
binaries \citep[see][for an explicit discussion about
these limits for extreme mass ratios]{Dominik1999}, which are illustrated in
\Fig{fig:caustopo}. 
In the close binary case (on the left in the figure), there exits a
central caustic near the more massive body, as well as two secondary 
caustics\footnote{In the case of planetary mass ratios ($q\ll1$),
  secondary caustics are usually referred as ``planetary'' caustics.}
above and below the binary symmetry axis.  
The intermediate binary configuration (in the middle) displays a
single caustic (also called a resonant caustic when $d\sim1$), while the
wide binary (on the right) involves
a central caustic plus an isolated single secondary caustic, located
between the two lens components.

It has been shown by \cite{ErdlSchneider1993} in the mathematical framework of
catastrophe theory that there exist close analytical formulae which
determine the bifurcation values between these regimes, which only
depend on the lens separation and mass ratio $(d,q)$:
\begin{equation}
  d_c^{\,8} = \frac{{(1+q)}^2}{27q}\,{(1-d_c^{\,4})}^3 \, ,
  \label{eq:bifurqc}
\end{equation}
\begin{equation}
  d_w^{\,2} = \frac{{(1+q^{1/3})}^3}{1+q} \, ,
  \label{eq:bifurqw}
\end{equation}
where $d_c(q)$ and $d_w(q)$ are respectively delimiting the
close-intermediate and intermediate-wide binary regimes for a given
value of the mass ratio $q$. The resulting frontiers are plotted in
\Fig{fig:caustopo}.

\subsection{The binary lens parameterisation} \label{sec:binlenspara}

\begin{figure}[!htpb]
  \begin{center}
   \includegraphics[width=8.5cm]{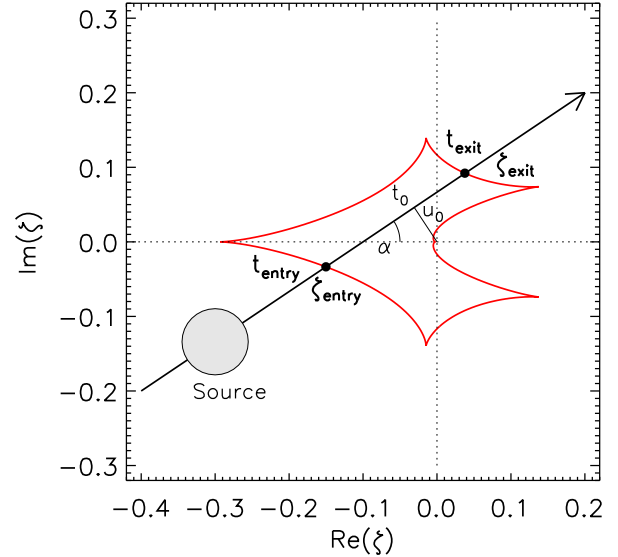}
    \caption{
      A binary lens generates  caustics, as
      displayed here (red lines) for a system of two masses
      located on the $x$-axis with separation $d=1$ and 
      mass ratio $q=10^{-2}$ (intermediate binary lens).
      The more massive (or primary) body is located
      at the origin of the coordinate system, while the lighter
      component is on the left. 
      In the simplest case, the source centre trajectory is
      classically described (in thin letters) by a straight 
      line that passes at time $\to$ at the minimum distance $\uo$ from the
      coordinate system, and makes an angle $\alpha$ with respect to
      the  $x$-axis. In the proposed alternative parameterisation
      (in bold face),  the
      source motion is expressed by means of the caustic entry and
      exit dates $\tent$ and $\texi$, which occur at positions 
      $\yent(\sent)$ and $\yexi(\sexi)$ on the caustics. The
      corresponding model parameters $\sent$ and $\sexi$ result from a
      proper definition of a curvilinear abscissa along the caustic
      lines (\cf \Sec{sec:abscissa}). 
    }
    \label{fig:trajet}
  \end{center}
\end{figure}

In the simplest case, modelling a binary-lens microlensing
event requires seven parameters, which are classically the lens separation
$d$ and mass ratio $q$, the time $\to$ of the closest approach of the
source to the origin (here the primary body), $\uo$ the corresponding
impact parameter in units of $\ThE$, $\tE$ the Einstein
crossing time, $\alpha$ the angle of the trajectory with respect to
the $x$-axis, and the source radius $\rhoS$ in units of $\ThE$.
The corresponding parameterisation is illustrated in \Fig{fig:trajet}.

While this straightforward parameterisation (hereafter referred to as
``classical'') is very well suited 
to describe single lens or non-caustic crossing binary lens events, it
suffers from being rather disconnected from observable features in
caustic-crossing light curves. The main reason is that the
corresponding parameters contain no information on where the source
crosses the caustics. 
As a consequence, widely exploring the classical parameter space in
caustic-crossing events poses an important computational challenge.
Firstly, because a large proportion of the tested models 
do not produce caustic crossings 
located at the observed position on the light curve, and only a 
fraction of the calculated models contribute efficiently in the
minimising numerical procedure to find the solution. Secondly,
caustic-crossing events are known to exhibit very uneven $\chi^2$
hyper-surfaces \citep[see such a $\chi^2$ map in e.g.][]{Kubas2005}, 
and usually harbour a large number of local minima which should be found 
and considered as competitive models to explain the light curve
\citep[e.g.][]{DominikBLampbi}. Furthermore, in 
standard minimising routines, parameters vary within bounded
intervals or around starting values, which may cause the routines to
become stuck in a false minima if one (or more) of the parameters is not
correctly optimised. Hence, parameterisations oriented towards observed
features in the light curve help to circumvent strong correlations and
parameter degeneracies. Additionally, a set of different starting
points should be considered to ensure the completeness of the
parameter space exploration. 

One of the requirements when building the new parameterisation is that
the number of parameters should remain the same as for the classical one, 
but with parameters which reflect as much
as possible the caustic crossing features that are seen in the light
curve. \cite{Albrow1999} proposed a parameterisation adapted to model a
light curve where one caustic crossing is clearly identified in the
light curve, and for which a straight line caustic approximation can
be applied. In the general context discussed in this section, we keep
to the more general case where finite source effects are calculated
with no approximation, because the new parameterisation only affects
the geometry of the problem and not the way to compute the
magnification (see \Sec{sec:elaboratemodels} for more details). 
Nevertheless, we examine in \Sec{sec:linecaus} the inclusion of the 
straight line caustic approximation in our framework, thus extending
the work initiated by \cite{Albrow1999}.

We therefore introduce in place of the four parameters $\uo$, $\alpha$,
$\tE$ and $\to$, the dates of the caustic crossing at the entry and
exit dates, $\tent$ and $\texi$, as well as the two corresponding curvilinear
abscissa $\sent$ and $\sexi$ along the caustic, so that
$\yent=\y(\sent)$ and $\yexi=\y(\sexi)$ are the points in the source
plane where the trajectory intersects the caustic (\cf \Fig{fig:trajet}). 
Within a suitable range for these four parameters,
a theoretical light curve will display caustic crossing
features around the observed entry and exit dates. 
The geometric conversion between the two set of parameters is
straightforward, since $\yent$ and $\yexi$ define the direction of the
source motion.

\subsection{Curvilinear abscissa along the caustic curve} \label{sec:abscissa}

In order to define a proper curvilinear abscissa, ones needs to build
an appropriate parameterisation of the caustic curves. A possible option
is to follow the method initiated by \cite{ErdlSchneider1993}, for
whom the caustics are described in a polar coordinate system $(r,\phi)$ in the
source plane. One then needs to solve for $\cos(\phi)$ for a given
value of $r$, hence parameterising the caustic line by $r$. A
working implementation of such a parameterisation is described in
detail by \cite{Dominik2007cont} in the context of image contouring.

In this paper, we work out an alternative method, based on a
result obtained by \cite{Witt1990}: for a point-source $\y$ exactly
located on a caustic, its corresponding point-like image $\x$ on the
critical curve in the lens plane satisfies the equation:
\begin{equation}
  \frac{1}{1+q}\left[\frac{1}{\x^2} +
    \frac{q}{(\x+d)^2}\right] = e^{-i\phi} \, ,
  \label{eq:Wittprim}
\end{equation}
where $\phi\in[0, 2\pi]$ is an {\it ad hoc} parameter, independent of
the choice of the coordinate system.
For a given binary lens $(d,q)$ configuration, we can compute for
every $\phi$ in the range $0-2\pi$ the caustic points by writing
\Eq{eq:Wittprim} as a complex polynomial of fourth degree: 
\begin{equation}
  \x^4 + 2\,d\,\x^3 + (d^2-e^{i\phi})\,\x^2 -
  \frac{2\,d\,e^{i\phi}}{1+q}\,\x - \frac{d^2\,e^{i\phi}}{1+q} = 0 \, .
  \label{eq:polyxy}
\end{equation}
For a given $\phi$, this equation has four distinct complex
solutions, which means the caustics are entirely built when $\phi$
runs from $0$ to $2\pi$. Moreover, since the caustics never intersect
in the case of a binary lens (contrary to a triple lens for
example), there exists a unique value of $\phi$ for a given $\y$ on
the caustic (otherwise one would have $\y(\phi_1)=\y(\phi_2)$ with
$\phi_1\neq\phi_2$, which would mean an intersection of the caustic
lines). In this respect, \Eq{eq:Wittprim} is optimal in the sense that
each solution of the equation maps one and only one point on the
caustic lines, and the caustics are fully drawn for $0\leq\phi<2\pi$.

\begin{figure*}[!htbp]
  \begin{center}
    \includegraphics[width=5.8cm]{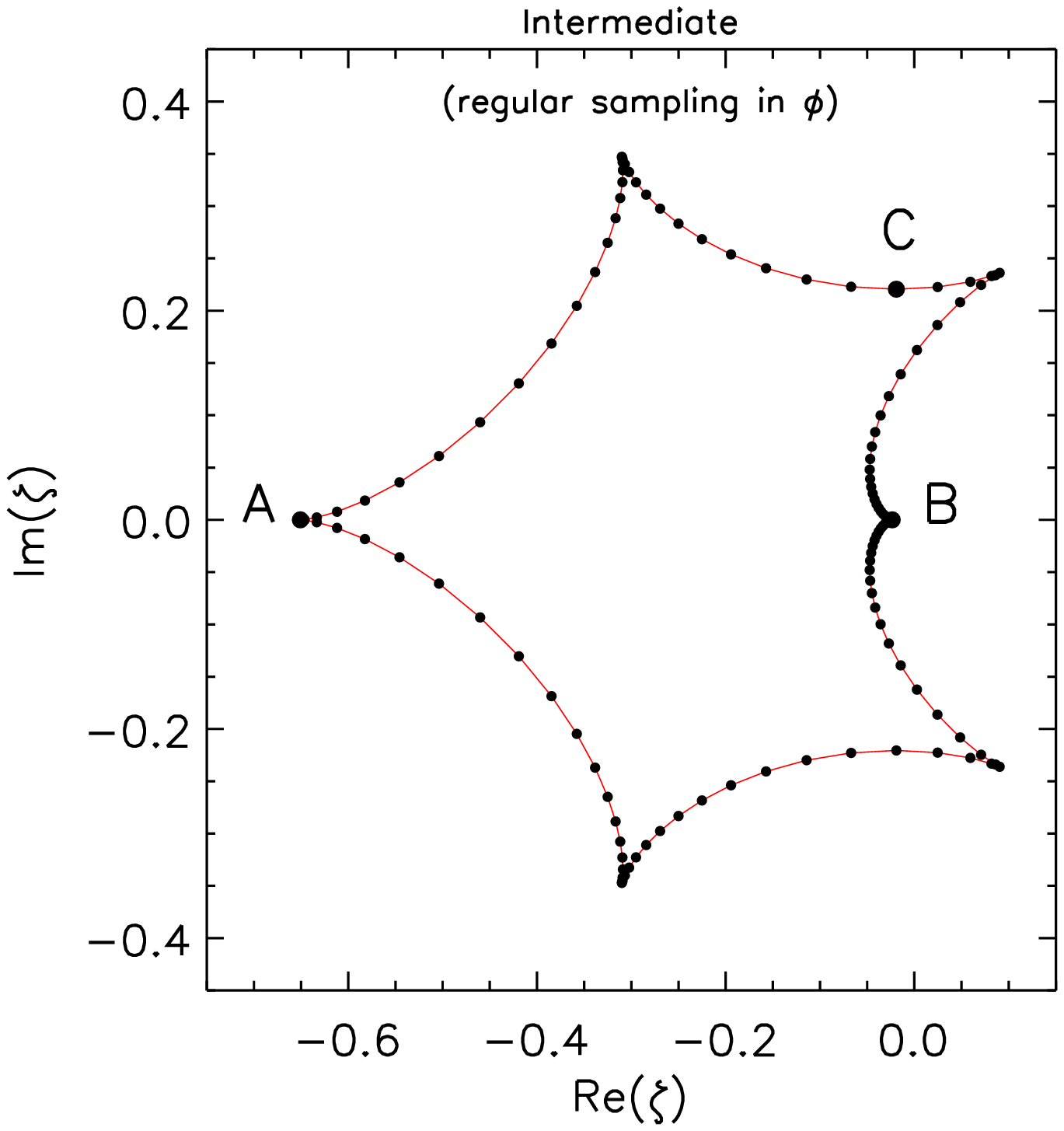}    
    \includegraphics[width=5.8cm]{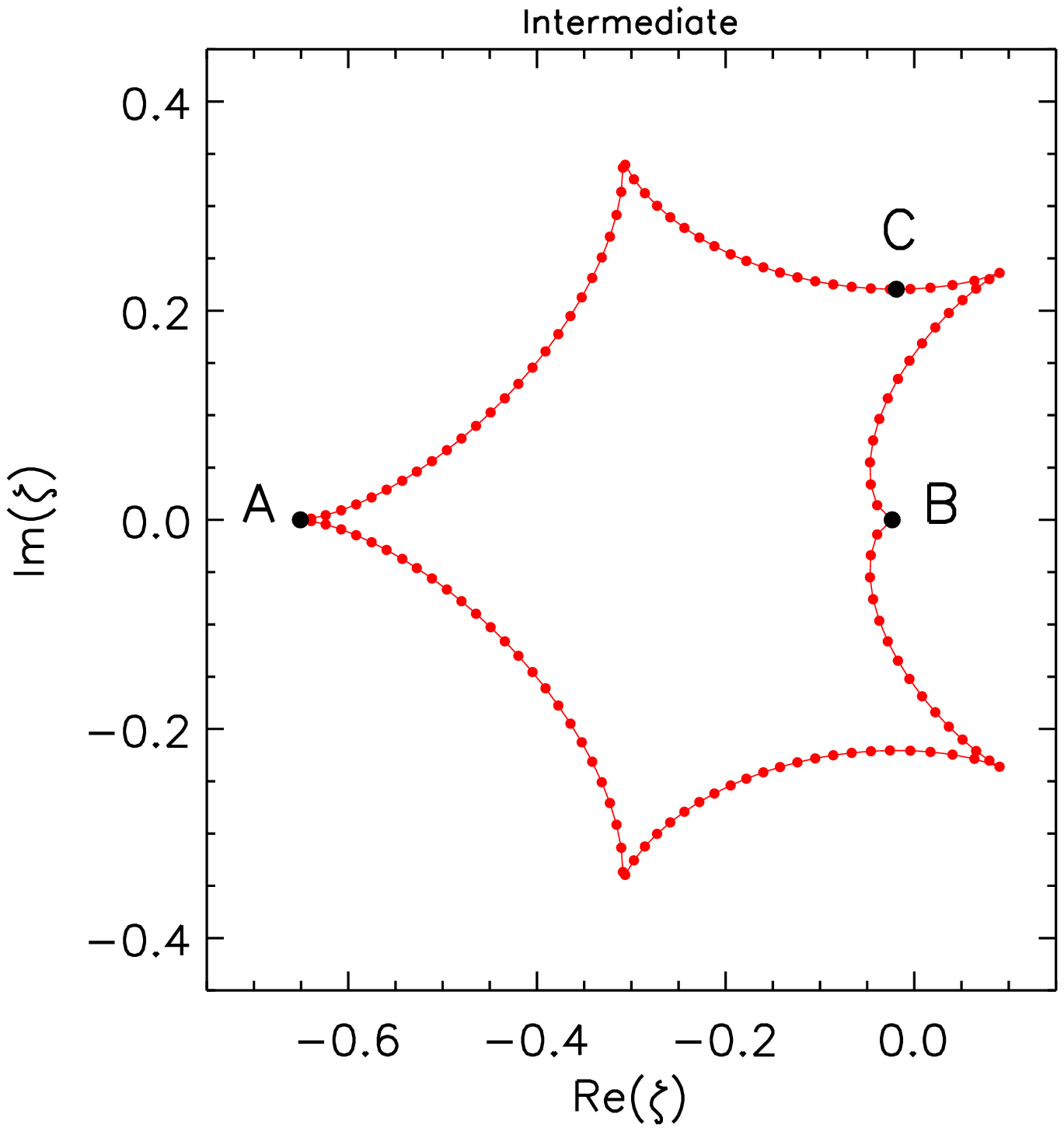}    
    \includegraphics[width=5.8cm]{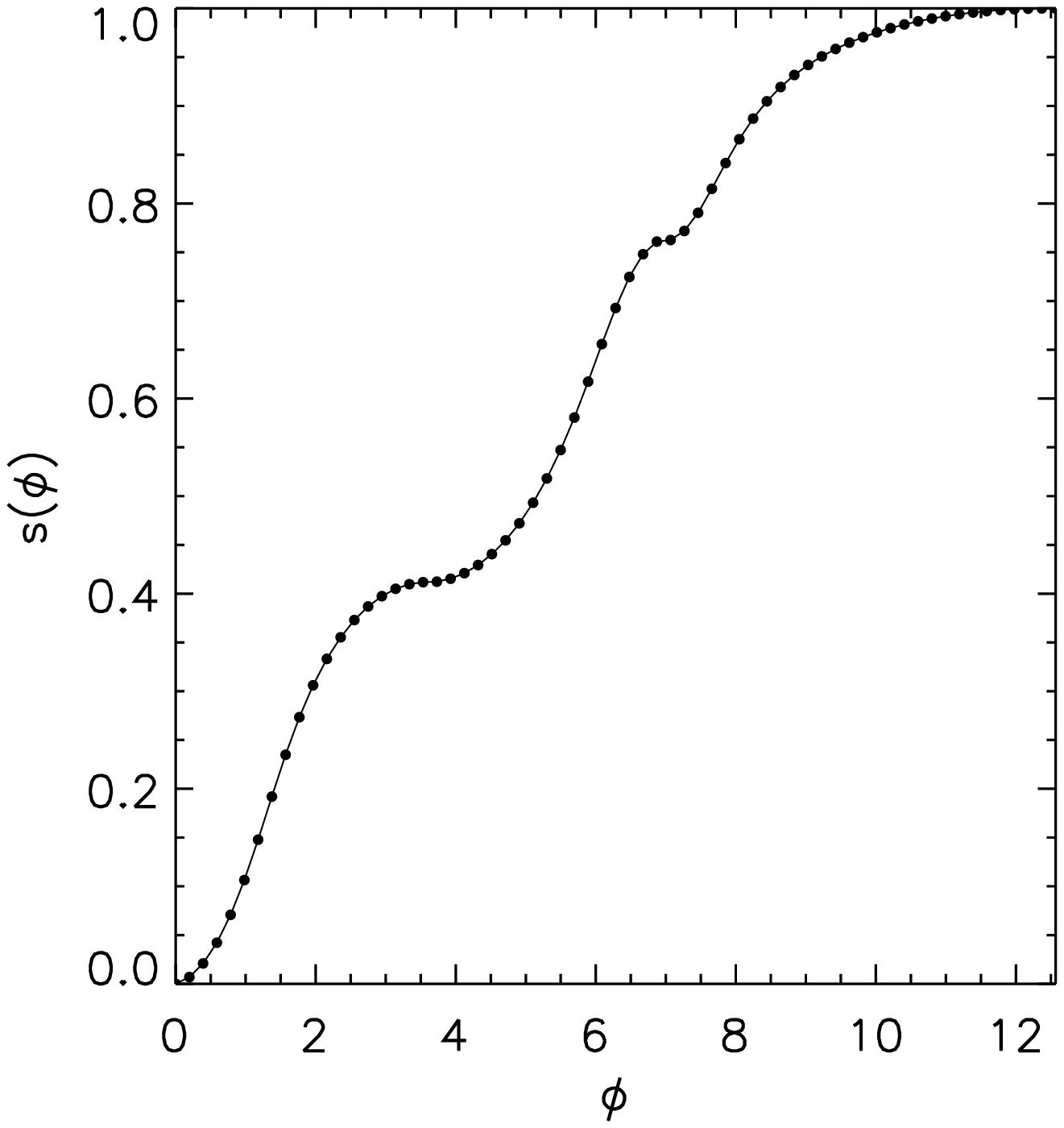}      \\
    \includegraphics[width=5.8cm]{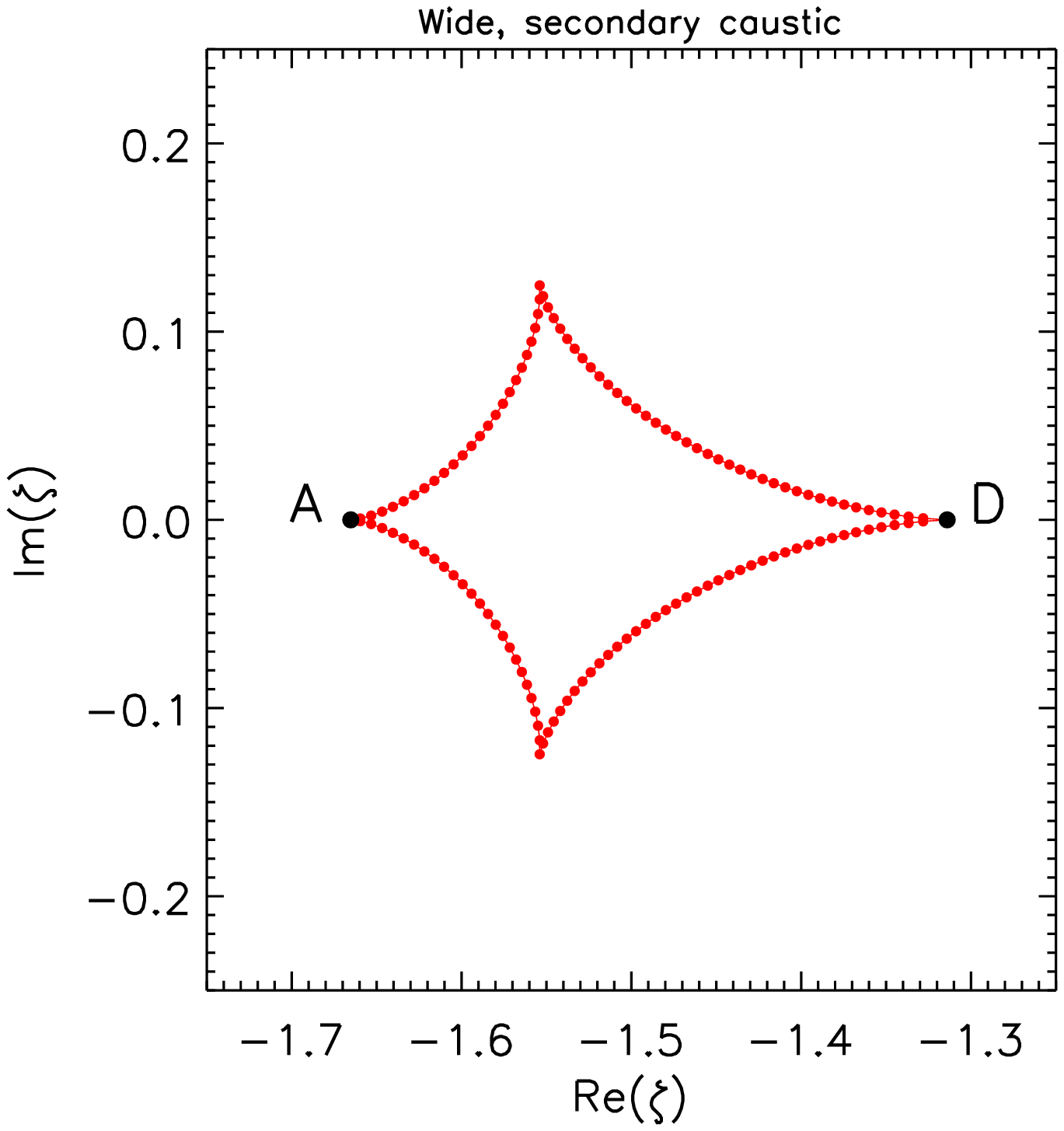} 
    \includegraphics[width=5.8cm]{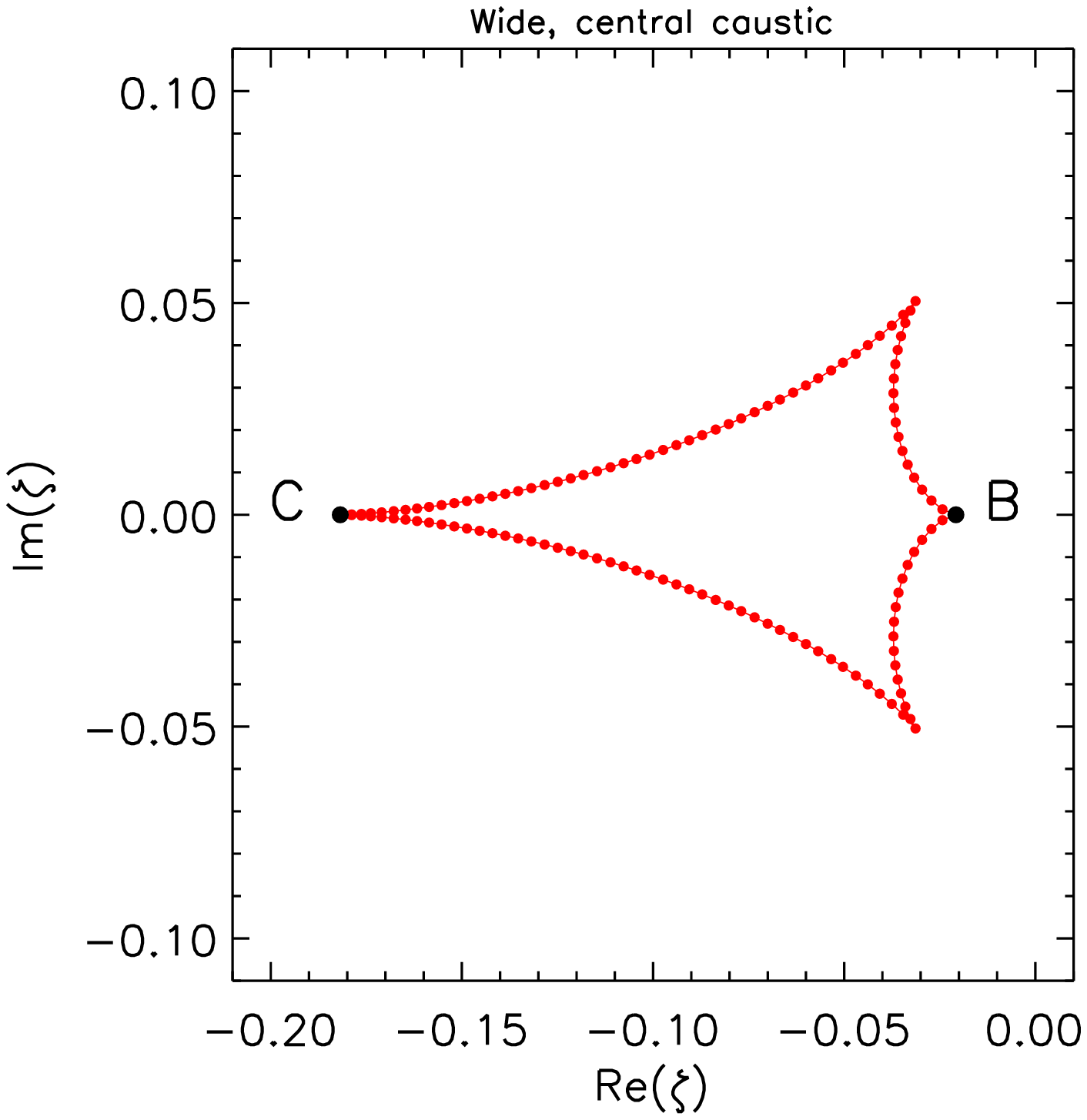}
    \includegraphics[width=5.8cm]{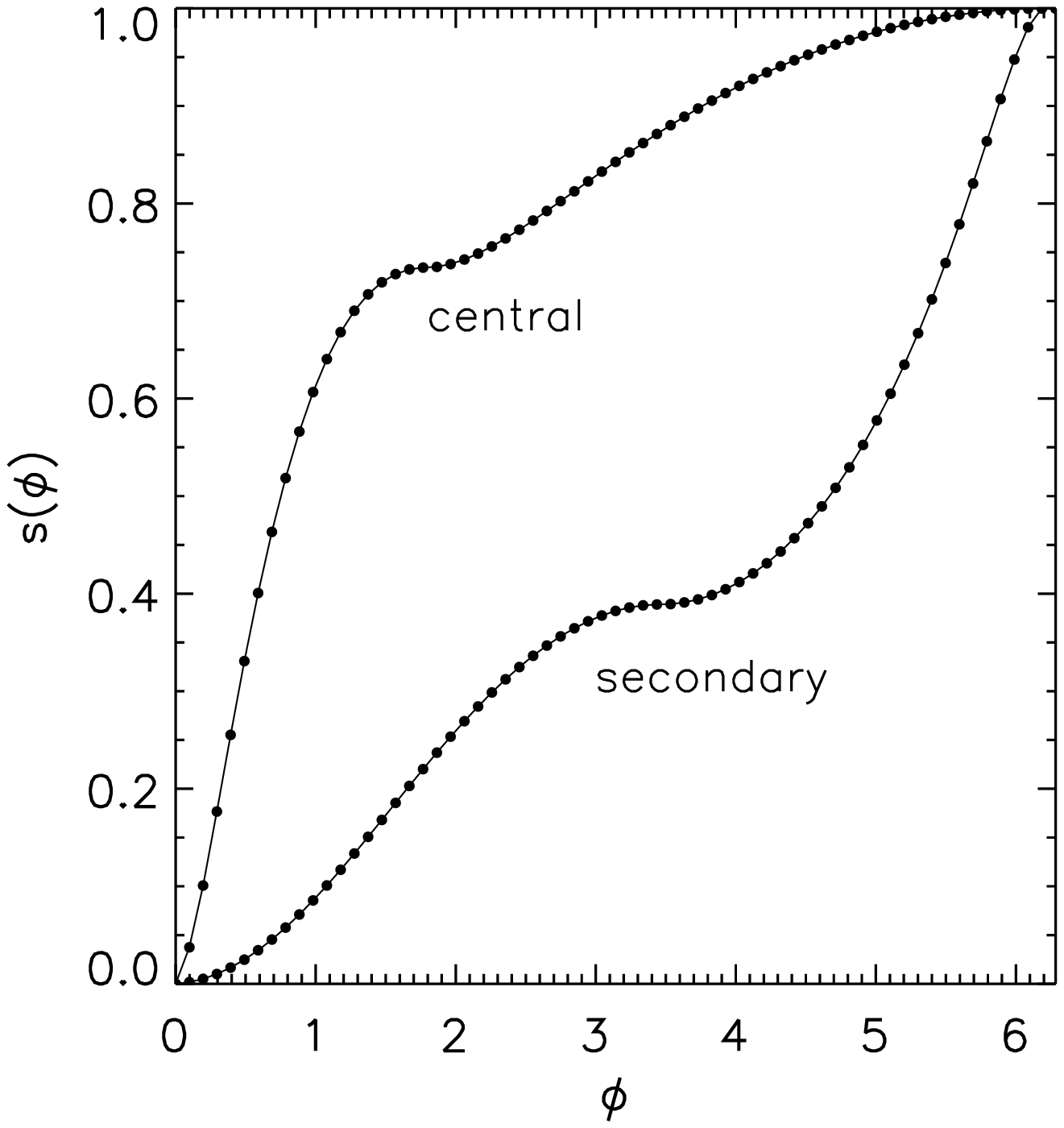}        \\
    \includegraphics[width=5.8cm]{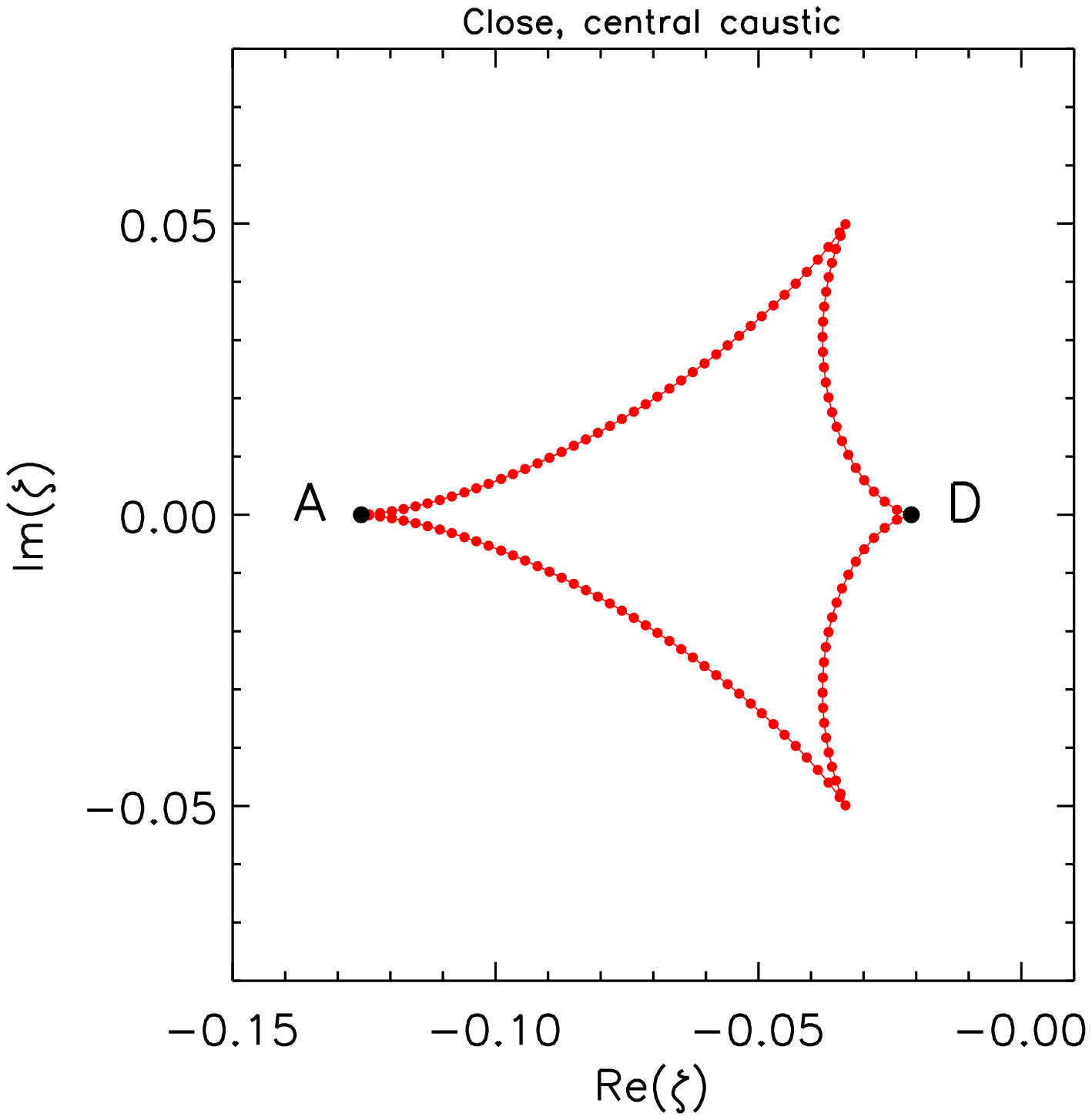} 
    \includegraphics[width=5.8cm]{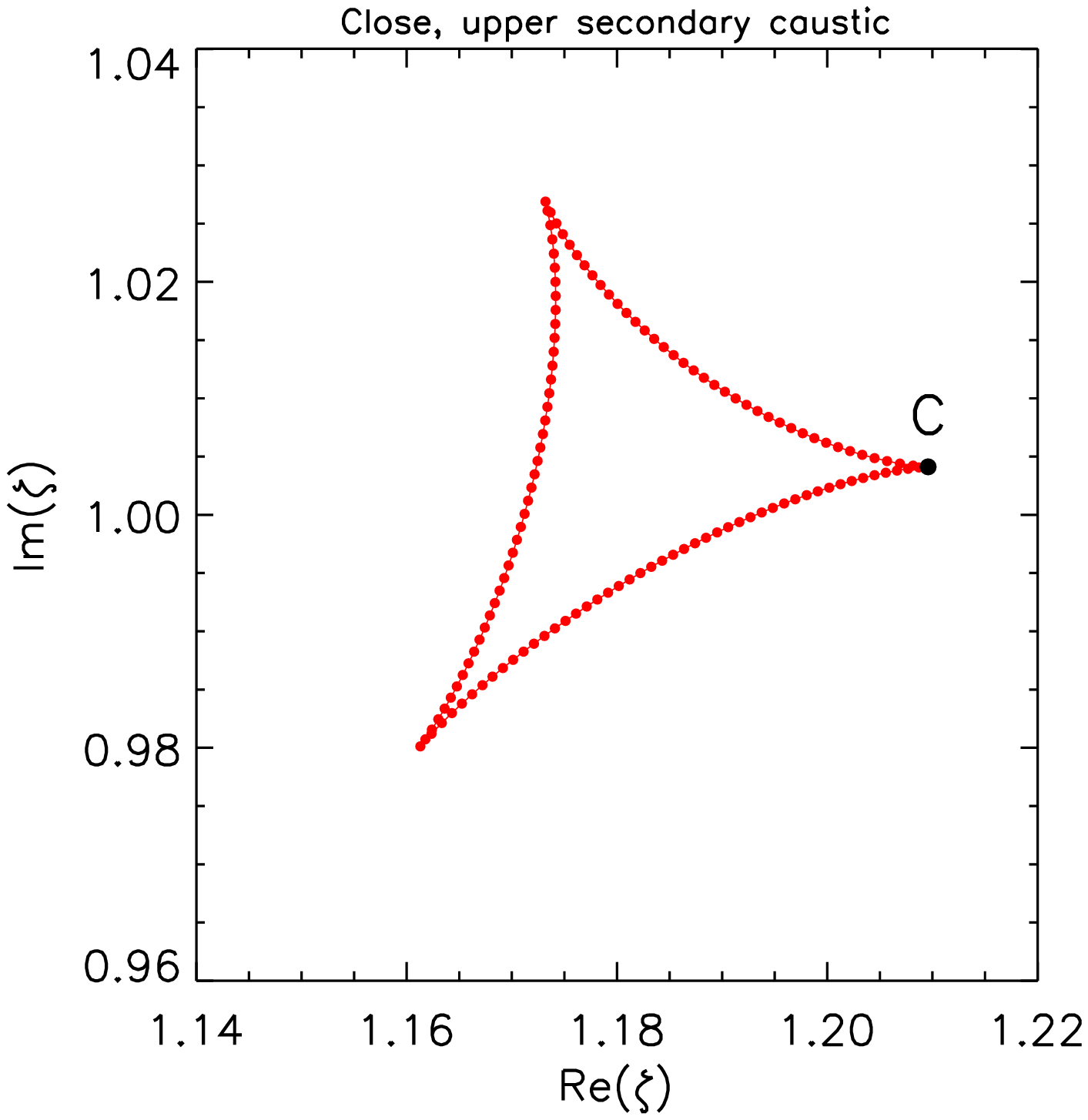} 
    \includegraphics[width=5.8cm]{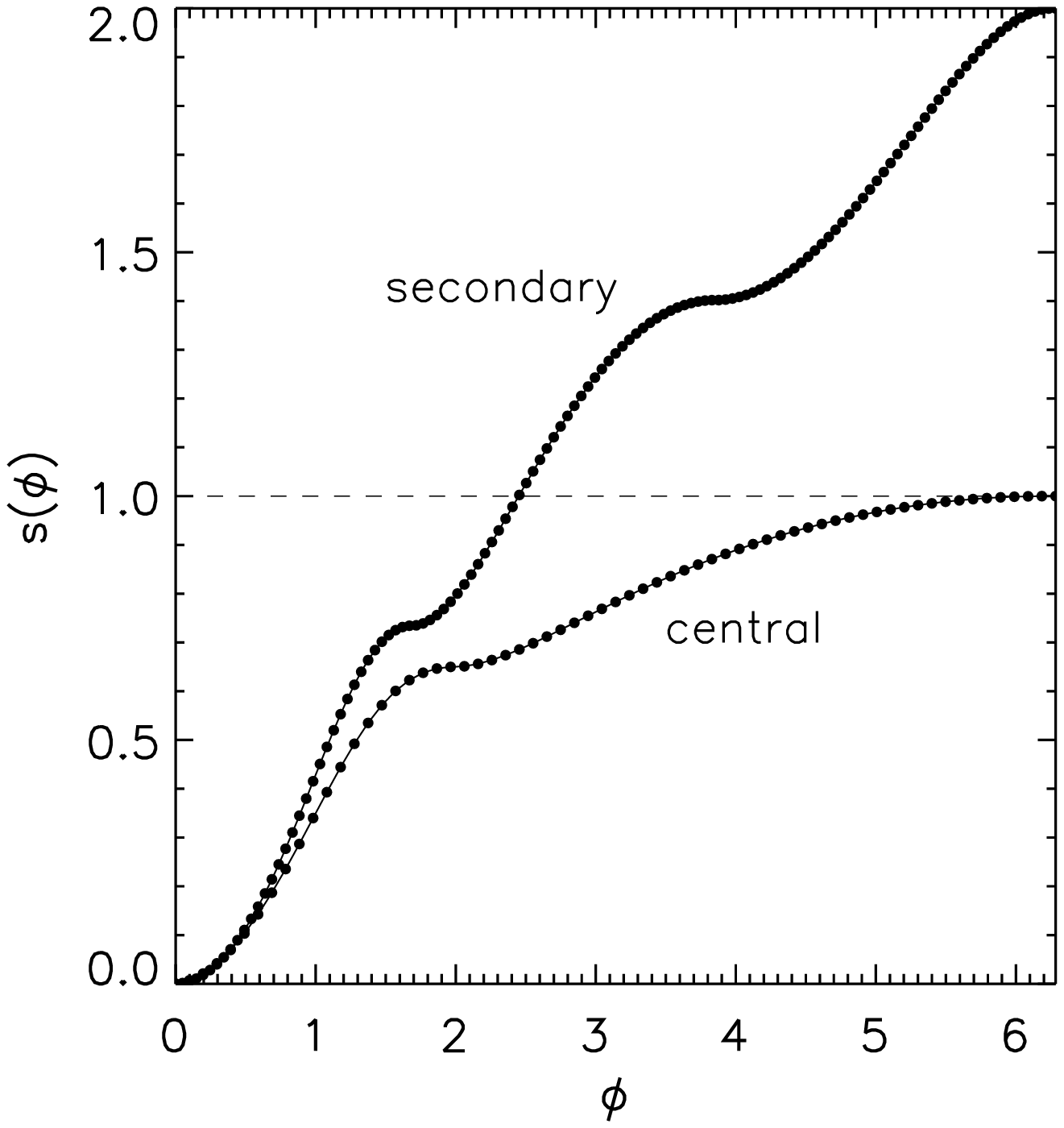} 
    \caption{
      The two left columns show the caustic
      patterns in the three possible topologies: intermediate (upper
      panels, $d=1.1$ and $q=0.1$), wide (middle panels, central and
      secondary caustics, $d=2$ and $q=0.1$) and close (central
      and upper secondary caustic, lower panels, $d=0.5$ and $q=0.1$).
      The primary lens is at the origin of the coordinate system, with
      the less massive body on the left.
      In the upper panel on the left, the sampled points on the caustic are
      computed assuming a regular spacing for $\phi$
      in equation \Eq{eq:Wittprim}, for a total of $130$ points (in
      practice this number is larger and is adjusted to match the desired
      precision in the sampling). 
      For comparison and using the same number of points, the next panel on
      the right shows the sampling of the caustics using a regular
      spacing for the curvilinear abscissa $s$ described in the paper,
      which results in a regular sampling of the caustics.
      The other plots display the wide and close caustics, also
      using a regular sampling for $s$.
      The three plots in the right most column show the function
      $s(\phi)$, which converts $\phi$ to $s$ according to
      \Eq{eq:defsc} and the choices made in \Sec{sec:numeric}.
    }
    \label{fig:causcurvi}
  \end{center}
\end{figure*}

We now examine how the caustics are drawn with $\phi$ for each of
the three topologies. Since the caustics are symmetric with respect to
the $x$-axis, we restrict ourselves to $\Im(\y) \geq 0$. By doing so, only two
of the four solutions of \Eq{eq:Wittprim} are taken into account. The
caustic region where $\Im(\y)<0$ is built by symmetry afterwards.
The corresponding caustics are drawn in \Fig{fig:causcurvi}, to which the
reader is referred in the following.
\begin{itemize}
\item In the intermediate binary case, $\phi=0$ leads to three
  points, A, B and C. When $\phi$ increases to $2\pi$, the
  caustics are simultaneously drawn by two distinct
  branches: A$\rightarrow$C and C$\rightarrow$B.
\item In the wide binary case, $\phi=0$ correspond to the four points
  A, B, C and D. When $\phi$ increases to $2\pi$, the central caustic
  is built with the single branch C$\rightarrow$B, while at the same
  time the secondary caustic is  built by the branch A$\rightarrow$D. 
\item In the close binary case, $\phi=0$ correspond to A, D and C.
  When $\phi$ increases to $2\pi$, the central caustic is built with
  the single branch A$\rightarrow$D, while at the same time the single
  branch  C$\rightarrow$C draws the {\it full} secondary upper caustic
  (so that no horizontal reflection is needed). The lower secondary
  caustic is then the symmetric image of the upper one.
\end{itemize}

A first naive choice would be to simply take $\phi$ as a curvilinear
abscissa. But, as shown in the upper panel on the left of
\Fig{fig:causcurvi}, a regular sampling on $\phi$ ({\it i.e.}
$\delta\phi=\K$) results in a very irregular linear sampling of the
caustics. Most of the points accumulate near the cusps, which makes
this parameter not well-suited. To avoid this effect, we define a
curvilinear abscissa $s$ which satisfies the property: 
$\delta s = \K \Rightarrow \left|\y(s+\delta
  s)-\y(s)\right| = \K^\prime$. 
Since at first order 
$\y(s+\delta s) \simeq \y(s) + \left(d\y/d s\right)\delta s$, the condition
to be fulfilled reads: $\left|d\y/d s\right| = K$, with $K$ a
constant, so that:
\begin{equation}
  \left|\frac{d\y}{d\phi}\right| = \left|\frac{d\y}{d
      s}\right|\times\left|\frac{d s}{d\phi}\right| = K
  \times\left|\frac{d s}{d\phi}\right| \, .
  \label{eq:dydphiK}
\end{equation}
The abscissa $s$ can finally be obtained as a function of $\phi$ by
integrating the derivative $d s/d\phi$, choosing
$s(\phi)$ monotonic:
\begin{equation}
 s(\phi) = \int_{0}^{\phi} \frac{d s}{d\phi}\,\d\phi + s(0) =
 \frac{1}{K}\,\int_{0}^{\phi}
 \left|\frac{d\y}{d\phi}\right|\,\d\phi + s(0) \, , 
 \label{eq:defsc}
\end{equation}
where the normalisation constant $K$ depends on the choice of the
ranges for $\phi$ and $s$. This is discussed in detail in
\Sec{sec:numeric} for the three topologies.
A remaining technical point here concerns the calculation of the
derivative $d\y/d\phi$. By differentiating
\Eq{eq:eqlensprim} with respect to $\phi$ and substituting
\Eq{eq:Wittprim}, we first obtain: 
\begin{equation}
  \frac{d\y}{d\phi} = \frac{d\x}{d\phi} +
  e^{i\phi}\,\frac{d\overline\x}{d\phi} \, ;
  \label{eq:dydphi}
\end{equation}
we then get the expression of $d\x/d\phi$ by writing
\begin{equation}
  \frac{d}{d\phi}e^{-i\phi} = \frac{d\x}{d\phi} \times
  \frac{d}{d\x}e^{-i\phi} = -i\,e^{-i\phi} \, ,
\end{equation}
which leads, after differentiating \Eq{eq:Wittprim} with respect 
to $\x$, to
\begin{equation}
  \frac{d\x}{d\phi} = \frac{i}{2}\,\frac{(\x+d)^2+q\,\x^2}
  {(\x+d)^3+q\,\x^3}\,(\x+d)\,\x \, ,
\end{equation}
and so $d\y/d\phi$ can be calculated analytically given a
point $\y$ on a caustic ({\it i.e.} a point $\x$ on a critical
curve). 

In \Fig{fig:causcurvi}, with the exception of the upper panel on the
left commented on previously, we have used a regular sampling of $s$ to
draw the caustic lines, following the above prescriptions. One can
notice the regular spacing of the points along the caustics, even
for the small number of sampling points chosen here ($130$ for a whole
caustic). 
When $\phi$ runs between $0$ and $2\pi$, it draws a caustic which may
contain cusp singularities. Because of the nature of such
a singularity, the tangential vector to the curve vanishes
\citep[e.g.][]{Witt1990}, 
and thus the derivative of $\y$ satisfies at the cusps locations:
$\left(d\y/d\phi\right)_{\rm cusp}=0$, or according to \Eq{eq:dydphiK},
$\left(d s/d\phi\right)_{\rm cusp}=0$. The cusps can be seen in 
\Fig{fig:causcurvi} at the location of the inflexion points of
$s(\phi)$.

\section{Implementation} \label{sec:numeric}

As seen in the previous section, the caustics are built by gathering
different branches obtained by moving $\phi$ from $0$ to $2\pi$. 
In our fitting strategy, we choose to independently parameterise the
central and secondary caustics, and make the choice that a complete
given caustic is fully drawn for $0 \leq s < 2$.

The abscissa $s$ is obtained by first
computing the caustics with a uniform sampling in $\phi\in[0, 2\pi]$,
and then computing numerically $s(\phi)$ using \Eq{eq:defsc}. The 
corresponding functions are plotted in the rightmost
column of \Fig{fig:causcurvi} for each of the three topologies.
As stated before, the cusps singularities can be seen as the inflexion
points of $s(\phi)$, because $\left(d\y/d s\right)_{\rm cusp}=0$. In
practice:
\begin{itemize}
\item In the intermediate binary case, since two branches are needed to
  build the upper part of the caustic, we choose $0 \leq \phi < 4\pi$,
  assuming that $0 \leq \phi < 2\pi$ builds the branch A$\rightarrow$C and
  $2\pi \leq \phi < 4\pi$ the branch C$\rightarrow$B. Since we have
  $0 \leq s < 1$ for half a caustic, we set $s(\phi=0)=0$ and
  $s(\phi=4\pi)=1$ in \Eq{eq:defsc}. The lower part of the caustic $1
  \leq s < 2$ is built by a horizontal reflection.
\item In the wide binary case, we separately treat the central and the
  secondary caustic. Since in both cases a single branch is needed to
  build the upper part of the caustic corresponding to $0 \leq s < 1$,
  we choose $s(\phi=0)=0$ and $s(\phi=2\pi)=1$. The lower part of the
  caustic $1 \leq s < 2$ is once again built by an horizontal reflection.
\item In the close binary case, we also separately parameterise the
  central and secondary caustics. The central caustic parameterisation
  is obtained in the same way as for the wide binary central caustic. 
  On the other hand, the upper secondary caustic is fully drawn with the
  single branch C$\rightarrow$C, so that we have $s(\phi=0)=0$ and
  $s(\phi=2\pi)=2$. The lower secondary caustic is just its horizontal
  reflection.
\end{itemize}  

Computing and parameterising the caustics is the most
time consuming part of the method.
Hence an efficient way to use this
parameterisation is to perform the fitting procedure
on a pre-defined $(d,q)$ grid. This strategy is
also strongly indicated when searching for all possible local minima
within a reasonable $\Delta\chi^2$ from the global minimum,
which can possibly lead to competitive solutions. For example, 
\cite{Kubas2005}  found two local minima of almost the same goodness of
fit for OGLE~2002-BLG-069L, one associated with a close binary
and a central caustic crossing, and the other involving a wide binary
and a secondary caustic crossing.

In practice, for a given $(d,q)$ grid point we determine to
which topology it corresponds (\cf \Eq{eq:bifurqc}) and
choose between the central and secondary caustics. A regular sampling
of the caustics is then performed, using a regular spacing for $s$.
The parameters $\tent$ and $\texi$ are then chosen so that during the
fitting process they stay close (\cf below) to the values guessed from
the light curve inspection.

\section{Fitting strategy} \label{sec:startegy}

\subsection{Extension to elaborated binary lens models} \label{sec:elaboratemodels}

A first important fact about the presented parameterisation is
that it is purely geometrical, and so does not affect the way the
magnification is computed. The magnification can be calculated via
classical methods, such as magnification maps
\citep{RattenburyRay,Wambsganss1997}, contouring methods
\citep{Dominik2007cont,GouldGaucherel1997}, improved ray shooting
techniques \citep{Dong2006} or special approximations of the
magnification function.
Thus, finite source effects, including
limb-darkening, can be used within this framework.

Secondly, since only the positions on the caustic and their
respective dates are fitted, one can consider a source
trajectory affected by effects like parallax 
\citep[see in particular][]{Gould2004JerkParallax} or xallarap 
\citep[motion of the source due to a massive but usually unseen
companion, e.g.][]{Jiang2004J,Ghosh2004}. One can also consider
binary lens rotation \citep[e.g.][]{Dong2008ob71}, by computing and
parameterising the $(d,q)$ configurations at the entry and the exit
dates, performing the appropriate coordinate rotation, and fitting 
e.g. for $(d_{\rm entry},d_{\rm exit},\tent,\texi,\sent,\sexi)$.

Last, one can efficiently use the parameterisation to model events
with poorly-sampled caustic-crossings \citep[see e.g.][]{Kains472}: it
prevents the code from converging towards a minimum that does
not exhibit the caustic-crossing features at the right place, which is
likely to happen with the classical parameterisation. Moreover it
allows us to inspect different scenarios with caustic-crossings occurring
at different possible estimated dates, such as for MOA~2007-BLG-197
\citep{Oasis} where the caustic entry date cannot be
directly inferred from the collected data.

\subsection{Fitting using standards minimising algorithms}

When the parameterisation is used with minimising algorithms like
the downhill simplex or conjugate gradient methods, which are 
well-suited to explore a small region around a local minimum, we
choose the starting points for $\tent$ and $\texi$ as 
estimated from the light curve inspection. 
The starting values for $\sent$ and $\sexi$
(both evolving in the interval $s\in[0,2]$),
can be randomly chosen, or be set to values that allow us to explore
trajectories of particular interest.
A drawback of using these minimising algorithms is they may fail to
locate all the minima, because they may converge to the nearest local
minimum. This can be overcome by setting different starting points for
$\sent$ and $\sexi$.
 
Genetic algorithms \citep[e.g.][]{Charbonneau1995} are quite
well-suited for use with the new parameterisation, because they combine
the possibility of restricting the exploration of $\tent$ and $\texi$ to
intervals compatible with the positions of the caustic crossings on
the light curve, along with a wide exploration of all kind of
trajectories by performing a dense sampling 
of $(\sent,\sexi) \in [0,2]\times[0,2]$. 
Because by essence a genetic algorithm does not converge to the exact
minimum in the best found $\chi^2$ valley, it is recommended to
further explore and refine the solution
by using one of the
method described above, or a Markov Chain Monte Carlo (MCMC, see
below). 

The new parameterisation appears also very convenient for MCMC
algorithms. The starting values for the parameters $\tent$ and
$\texi$ can be estimated from the light curve as already mentioned,
while the one-$\sigma$ intervals of variation around them can be
adjusted to their estimated error. One can then perform a wide
exploration of all possible combinations of caustic crossing positions 
$\sent$ or $\sexi$, or refine the parameters around some already found
local minima, thus choosing a small corresponding $\sigma$.
We find in particular that the new
parameterisation greatly increases the final mean acceptance rate. MCMC
also has the great advantage of providing 
the error bars on the best set of parameters without further calculations.

Finally, an interesting property of the parameterisation
is that the event time scale, $\tE$, is computed rather than fitted as
the usual case, so that
\begin{equation}
  \tE = \frac{\texi-\tent}{|\yent-\yexi|}
  \label{eq:backtE}
\end{equation}
for a straight-line and uniform source trajectory.
This formula shows that there is a strong correlation between
the distance between the two caustic crossings and $\tE$.
Therefore it can happen that the minimising algorithm is not able to
explore adequate values for $\tE$, because for example the starting point
was chosen too far from its final optimal value. It is then very
likely to stop at a false minimum. Far from being a casual case,
this is very likely to happen for binary lens caustic crossing
events, for caustics show some similarity in
their shape and magnification pattern over a wide range of binary
separations and mass ratios 
\citep[see e.g. figures in][]{Han2006planCaus,Chung2005,Dominik1999},
which may draw a comparable magnification light curves, but for a
range of values for $\tE$. Using the new parameterisation,
there is no such problem because $\tE$ is not constrained, which is
one of its advantages. 
Finally, since the best models should be associated with acceptable
physical parameters, the values of $\tE$ found for the best model(s)
can be used as a consistency check of their physical plausibility.

\section{The straight line caustic approximation} \label{sec:linecaus}

\subsection{Assumptions and application}

 The magnification in the vicinity of fold caustics
 has been studied in detail by a number of authors 
 \cite[see in particular][]
 {SchneiderWeiss1986,Gaudicaustics2002,Dominik2004PracticeFold},
 and one of the most important results in practice is that far
 enough from cusp singularities, the point-source magnification
 follows a generic behaviour. In particular, when a point-source 
 crosses a caustic, a pair of critical images are created or merge at
 the critical line in the lens plane,
 depending on the source direction. At first order, their
 magnification is the same, and is solely dependant on the point-source
 distance $\yperp$ from the caustic, 
 with $A \propto K/\sqrt{\yperp}$, where the coefficient $K$ is
 related to the local properties of the lens equation.

 A first condition for a caustic to be treated as a straight line is
 that the local curvature radius 
 of the caustic\footnote{The local curvature of the caustic 
   can be worked out analytically using in particular \Eq{eq:dydphi}.} 
 should be much greater than the source radius.
 Moreover, the gradient of the magnification along the caustic should 
 also be negligible, which is achieved when the source crosses the
 caustic far enough from a cusp. Within these assumptions,
 semi-analytical formulae can be derived which allow a very efficient
 computation of the time-dependant magnification during the caustic
 crossing, instead of the very demanding full finite-source integration.

 There are many possible applications of this approximation. 
 In order to study microlensing effects in quasar lensing, for example,
 \cite{Yonehara2001} makes use of a straight line fold caustic
 approximation to model a suspected caustic crossing 
 affecting image ``C'' of quasar Q2237+0305.
 In the context of stellar microlensing, \cite{Albrow1999} have worked
 out an improved parameterisation for binary lenses involving one
 well-covered caustic crossing. In this framework, the new set of parameters
 is directly linked to the properties of the observed caustic crossing,
 such as the source radius caustic-crossing time, or the date at which
 it happens. This method was successfully applied to a binary lens event
 in the Small Magellanic Cloud microlensing, MACHO~98-SMC-1
 \citep{Afonso2000}. Caustic crossings are also well-suited to
 study source stars' brightness profiles, because different
 regions across their disks are differentially magnified:
 For example \cite{Cassan69letter} used the
 straight line caustic approximation to study the source
 limb-darkening for OGLE~2002-BLG-069
 without requiring any knowledge of the full binary lens
 model. This is of great interest  when the full light curve is not
 well covered, or when many different binary lens models fit the light
 curve. If the lens is a strong binary, with typical mass ratio
 $q \sim 0.1 - 1$, the caustics have a large spatial extension,
 which gives the method a high potential in practice.

 As mentioned in \Sec{sec:binlenspara}, we now examine in
 detail how the straight line caustic approximation can be introduced
 in our framework, hence providing an extension of the work initiated
 by \cite{Albrow1999}.
 In the following, we consider a given and fixed binary lens
 configuration $(d,q)$.
 While outside of a caustic region a point-source is mapped onto three
 ``non-critical'' images, it is well known that a further pair of 
 ``critical'' point-like images appear when the source crosses the 
 caustic \citep{BourassaKantowski1975,SchneiderWeiss1986}. The
 magnification for an individual image $\x$ issued from a
 point-source located at $\y$ is given by   
 \begin{equation}
   A^{\rm PS}(\x) = \left|\frac{1}{\det\vec{I}}\right| \, ,
   \label{eq:A_PS}
 \end{equation}
 where $\vec{I} = \partial(\xi,\eta)/\partial(x,y)$ is the Jacobian
 matrix associated with the lens equation \Eq{eq:eqlensprim}, with
 $\x=x+iy$ and $\y=\xi+i\eta$.

 Under the reasonable assumption that for many cases the source
 size is small with respect to the caustics,
 the three non-critical images
 are weakly affected by finite source effects during a caustic crossing.
 On the other hand, the
 magnification of the two critical images is strongly dependant on the
 finite extension of the source. We thus divide the total magnification
 into two contributions,
 \begin{equation}
   A\left(\y, \rhoS \right) = \Anc\left(\y, \rhoS \right) + \Ac\left(\y, \rhoS\right) \, ,
   \label{eq:Atot}
 \end{equation}
 where $\Anc$ stands for the three non-critical images total
 magnification and $\Ac$ for the two critical images total
 magnification, with $\rhoS$ the source radius in units of $\ThE$.
 We detail in the following sub-sections
 how we can treat $\Anc$ and $\Ac$ in practice.

\subsection{The non-critical images magnification}

In previous studies making use of the straight line caustic
approximation, the total magnification of the non-critical images
is approximated at first order by the
Taylor expansion $\Anc \simeq a + \omega\,(t-\tc)$, where $\tc$ is the time
of the caustic crossing (in the following, we use the subscript
``$c$'' for parameters computed on the caustic or critical lines,
corresponding to the ``entry'' or ``exit'' used throughout the paper).
Within this scheme, $a$ and $\omega$ are fitted
parameters. This choice is justified in alternative parameterisations
such as in \cite{Albrow1999}, where only one caustic crossing is
studied, or in \cite{Cassan69letter} where the caustic crossing is
modelled alone, assuming no knowledge of the full binary lens  
model.
In the parameterisation presented here, it would imply adding two more
parameters per caustic crossing, for a total of four more parameters
in addition to $\tent$, $\texi$, $\sent$, $\sexi$,
while the classical full binary lens parameterisation involves only one
more parameter to treat finite source effects, $\rhoS$ (assuming a
uniformly bright source). Moreover, it can happen that for a caustic
crossing located at the top of a microlensing light curve, the linear
approximation of $\Anc$ is not precise enough because of a somehow
stronger local curvature in the time-dependant magnification.

Here we proceed in a different way: we choose to directly identify and
track the three non-critical images during the whole caustic
crossing, in order to compute their individual magnification.
This can be done in the following way: at the position
$\yc$ where the source centre crosses the caustic line, we use the lens
equation \Eq{eq:eqlensprim} to locate the five images; the two
critical images appear at the same position $\xc$ on the critical
curve in the lens plane, with $\xc$ and $\yc$ satisfying 
\Eq{eq:Wittprim} as well. This provides a very robust
numerical test to locate the two critical images --  and therefore
the three non-critical images at the caustic crossing date. A
Newton-Raphson method can then be implemented to ``follow'' the
non-critical images and compute their point-source magnification
\begin{equation} 
  \Anc(\y,\rhoS \equiv 0) \simeq \sum\limits_{i=1}^{3}\,A^{\rm PS}(\x_i) 
\end{equation}
at any position during the caustic crossing. 
Alternatively and for more precision, one may replace the point-source
approximation by a Taylor expansion of the lens equation, and perform
its integration over the source disk. 
Such an expansion of the magnification up to $O(\rho^6)$
has been studied in the general case by \cite{PejchaHeyrovsky2008}, 
while a numerical implementation up to the hexadecapole for
non-critical images has been presented in detail by \cite{GouldHDP}.
The derived formulae do not require a further numerical integration,
thus allowing a very efficient computation 
of $\Anc(\y, \rhoS \nequiv 0)$, which is one
of the main requirements about the method outlined in this section.
In addition to this, \cite{GouldHDP} demonstrates that after few source
radii from the caustic, the hexadecapole approximation is
of very high precision.

\subsection{The critical images magnification}

For the critical images, the only relevant quantity in the
straight line caustic crossing approximation 
is $\yperp$, the perpendicular distance
from a point-source $\y$ to the caustic.
Following our choice of curvilinear abscissa along the
caustic (\cf \Sec{sec:numeric}), the inside 
region delimited by the caustic curves is always on the right
hand side of the tangential vector, so that $\yperp > 0$ inside.
As shown in particular by \cite{SchneiderWeiss1986},
the magnification of a point-like critical image in the
vicinity of a caustic (and far enough from a cusp singularity) can be
approximated to the first non-vanishing order by:
\begin{eqnarray}
  A_{\rm C}^{\rm PS}(\yperp > 0) &\simeq& \frac{1}{2}
  \left|\frac{\lambda}{\mu_{1}\sqrt{2\,(1-\lambda)}}\right|^{1/2} 
  \frac{1}{\sqrt{\yperp}} \, , \\
  A_{\rm C}^{\rm PS}(\yperp \leq 0) &=& 0 \, ,
  \label{eq:Ac1}
\end{eqnarray}
where $\lambda = 1 - \partial \xi/\partial x$ and 
$\mu_{1} = \partial\lambda/\partial y$ are evaluated 
at $\x = \x_{c}$.
Within these assumptions, the two critical point-like images are of
equal magnification; the corresponding total finite source
magnification is thus obtained by integrating 
$2 \times A_{\rm C}^{\rm PS}$ over the source disk.
Assuming a uniformly bright source disk,
and following \cite{SchneiderWeiss1987}, one obtains:
\begin{equation}
  \Ac(\y, \rhoS) \simeq
  \left|\frac{\lambda}{\mu_{1}\sqrt{2\,(1-\lambda)}}\right|^{1/2}
  \rhoS^{-1/2}\,J\left(\frac{\left|\y-\yc\right|
      \sin\psi_c}{\rhoS}\right)
  \label{eq:Jd}
\end{equation}
where $J$ is the semi-analytical function defined in Eq.~(B.7) of
\cite{SchneiderWeiss1987} and $\y$ the centre of the source;
$\psi_c$ is the angle between the source trajectory and
the tangent to the caustic at critical point $\yc$.
In practice, $\psi_c = \alpha - \gamma_c$, where $\alpha$ is the
trajectory angle of \Fig{fig:trajet}, and $\gamma_c$ is the angle
between the $x$-axis and the tangent to the caustic. 
Since the latter is collinear to $(d\y/d\phi)_\xc$, it can be
evaluated analytically using \Eq{eq:dydphi}.
A linear limb-darkened source model can
also be implemented very efficiently by adding one more parameter per
filter and computing further basis functions like $J$
\citep{Albrow1999,CassanPhDT,Cassan254}.

In practice, the function $J$ can be easily tabulated 
to the required precision,
for it only depends on the ratio $\upsilon=\yperp/\rhoS$. This is a key
aspect of using the straight line caustic approximation, for one
can replace a very demanding full finite-source computation by a
straightforward calculation of \Eq{eq:Jd}, thus speeding up the computing
time by many orders of magnitude. The tabulation of $J$ can be
restricted to a small interval, $\upsilon=0$ to few units (see below).

\subsection{The non-caustic crossing magnification}

Outside of the caustic crossing, the point-source or a
finite-source approximation like the hexadecapole
is used, while their validity for a given accuracy has to be estimated.
In the simplest case of the point-source approximation, for example,
the finite-source magnification of a critical image rapidly
converges toward the classical point-source magnification:
the absolute difference between $A^{\rm PS}$ and $\Ac$ is
less than $\sim 6\dix{-3}$ when the source centre is farther than
three source radii from the caustic,
and $\sim3\dix{-4}$ for ten radii. 

One way to ensure that the source crosses the caustic far enough from
a cusp is to locate the curvilinear abscissa $s_c$ as done in
\Sec{sec:abscissa} and to prevent the code from exploring values of $s$
in the vicinity of $s_c$. However if by chance the crossing occurs too
close to a cusp, we expect a rather poor agreement of the
magnification at the connection between the caustic crossing and
non-caustic regions, which would naturally exclude the configuration
obtained and thus minimise the effect of an inappropriate use of the
approximation. 

Finally, within the straight line caustic approximation for both the caustic
entry and exit, only one parameter remains to be fitted to
account for finite source effects: the source radius $\rhoS$, as
in the classical parameterisation.

\section{Summary and prospects} \label{sec:conclusion}

 We have presented an alternative parameterisation of the caustic
 crossing binary lens problem, which has the main advantage 
 of being directly linked to observed features in the light curve. We have
 explicitly discussed the underlying concepts and derived the
 corresponding equations, suggested a practical implementation, as
 well as examined its implications in terms of numerical
 robustness. Furthermore, we have shown that the straight fold caustic
 approximation is particularly easy to include in the presented
 framework, thus allowing a very high computation efficiency under
 some assumptions. 

 Current survey and follow-up networks of telescopes gather
 microlensing data at an accelerating rate, making the modelling a
 rather challenging task. Moreover, observations now involve new
 generations of automated telescopes, such as the RoboNet
 \citep{Robonet2007} and RoboNet-II \citep{Tsapras2008rn} initiatives, 
 which will have in the near future the
 power to track many more microlensing events. In this respect, a more
 automated modelling tool would be helpful to catch up with
 the flow of incoming data. Indeed, such a requirement for more automated
 software also applies for telescopes based in isolated or remote
 locations, like Dome~C in Antarctica \citep[see for example prospects
 underlined in][]{BeaulieuCassan2005DomeC}, 
 where microlensing would benefit from a continuous
 monitoring of the Galactic bulge in addition to the site's climatic
 advantages. 

 The automation of algorithms dedicated to microlensing event
 prioritising has seen important efforts
 \citep{Signalmen2007,Snodgrass2008plop}, but similar progresses in 
 modelling tasks remain to be achieved on the mathematical and
 numerical sides to keep up with the increasing rate of detection
 of anomalous microlensing events, which is one of the
 motivations of this work. 

\begin{acknowledgements}
  I would like to thank N.~Kains, D.~Kubas, C.~Snodgrass and
  Y.~Tsapras for very valuable comments on
  the manuscript, as well as the anonymous referee whose comments
  and suggestions helped improve the present work. 
  I also thank A.~Yonehara and R.W.~Schmidt for inspiring
  discussions on caustic topologies and properties.
  I thank K.~Horne for inviting me for a visit at SUPA, University
  of St. Andrews, where this work was initiated. 
  I acknowledge the ESO Director General Discretionary Found (DGDF) for
  financing a one-month visitor stay at ESO Santiago in March 2008,
  as well as S.~Brillant \& D.~Kubas for the invitation.
\end{acknowledgements}

\bibliographystyle{aa}
\bibliography{9795}
\end{document}